# Experimental demonstration of Luneburg waveguides

Vera N. Smolyaninova,[1] David Lahneman,[1] Todd Adams,[1]  Thomas Gresock,[1]

Kathryn Zander,[1] Christopher Jensen,[1] and Igor  I. Smolyaninov [2]

[1] *Department of Physics Astronomy and Geosciences, Towson University,*

*8000 York Rd., Towson, MD 21252, USA*

[2] *Department of Electrical and Computer Engineering, University of Maryland,*

*College Park, MD 20742, USA*

**Transformation optics gives rise to numerous unusual optical devices, such as novel metamaterial lenses and invisibility cloaks. Very recently Mattheakis *et al.* [1] have suggested theoretical design of an optical waveguide based on a network of Luneburg lenses, which may be useful in sensing and nonlinear optics applications. Here we report the first experimental realization of such Luneburg waveguides. We have studied wavelength and polarization dependent performance of the waveguides.**

Explosive development of elecromagnetic metamaterials and transformation optics (TO) produced such novel and fascinating optical devices as perfect lenses [2], hyperlenses [3-5], invisibility cloaks [6-9], and perfect absorbers [10]. Very recently Mattheakis *et al.* [1] have suggested a theoretical design of an optical waveguide based on a network of TO-based lenses, such as a Luneburg lens [11], and suggested that such a waveguide may be useful in sensing and nonlinear optics applications. Here we report the first experimental realization of such Luneburg waveguides, which operate in the visible frequency range. The individual Luneburg lenses in the fabricated waveguides are based on lithographically defined metal/dielectric waveguides [12-14]. Adiabatic variations of the waveguide shape enable control of the effective refractive index experienced by the TM light propagating inside the waveguide. Our experimental designs appear to be broadband, which has been verified in the 480-633 nm range. These novel optical waveguides considerably extend our ability to control light on sub-micrometer scales.

We have recently demonstrated that metamaterial parameter distribution required for TO-based designs can be emulated by adiabatic changes of shape of a 2D metal-dielectric optical waveguide [9,14]. Devices employed in our experiments have a three-layer waveguide geometry which is shown schematically in the inset in Fig.1(a). Assuming adiabatic changes of the waveguide thickness, the wave vector $k$ of the guided mode can be calculated as a function of light frequency $\omega$ and waveguide thickness $d$ for TE and TM polarized light, resulting in the definition of the effective refractive index $n_{eff}=k\omega/c$ for both polarizations [14]:

$$\left(\frac{k_1}{\varepsilon_m}-\frac{ik_2}{\varepsilon}\right)\left(k_3-\frac{ik_2}{\varepsilon}\right)e^{-ik_2d}=\left(\frac{k_1}{\varepsilon_m}+\frac{ik_2}{\varepsilon}\right)\left(k_3+\frac{ik_2}{\varepsilon}\right)e^{ik_2d} \qquad (1)$$

for the TM, and

$$\left(k_1-ik_2\right)\left(k_3-ik_2\right)e^{-ik_2d}=\left(k_1+ik_2\right)\left(k_3+ik_2\right)e^{ik_2d} \qquad (2)$$

for the TE polarized guided modes, where the vertical components of the wave vector $k_i$ are defined as:

$$k_1=\left(k^2-\varepsilon_m\frac{\omega^2}{c^2}\right)^{1/2}, \; k_2=\left(\frac{\omega^2}{c^2}\varepsilon-k^2\right)^{1/2}, \text{ and } k_3=\left(k^2-\frac{\omega^2}{c^2}\right)^{1/2} \qquad (3)$$

in metal, dielectric, and air, respectively. In the limit $\varepsilon_m \rightarrow -\infty$ Eqs. (1,2) are simplified as follows:

$$k_3 \cos k_2 d = k_2 \sin k_2 d \qquad \text{for TM,} \quad \text{and}$$

$$k_3 \sin k_2 d = -k_2 \cos k_2 d \qquad \text{for TE} \qquad (4)$$

Solutions of Eqs.(4) are plotted in Fig. 1(a), which shows the resulting effective refractive indices for both polarizations. This behavior may be used to fabricate various TO-based lenses if a waveguide thickness as a function of spatial coordinates $d(r)$ may be controlled with enough precision. For example, a modified Luneburg lens [11] with radial refractive index distribution

$$n = \sqrt{1 + f^2 - (r/a)^2} \, / \, f \quad \text{for } r < a \, , \qquad (5)$$

in which the refractive index varies from $n(0) = \sqrt{1 + f^2} \, / \, f$ to $n(a)=1$ is easy to realize for TM polarized light based on the comparison of experimental and theoretical data plotted in Fig.1(b,c). Theoretical performance of such a lens for $f = 1$ is presented in Fig. 1(d) based on the COMSOL Multiphysics simulations. On the other hand, the same $d(r)$ profile produces a different refractive index distribution for TE polarized light, which changes from $n(0) \sim 1.41$ to $n(a) \sim 0$. Due to near zero effective refractive index near the lens edge, the same device will operate as a spatial (directional) filter for TE light, as shown in Fig.2(b). This result is natural since most of TE light must experience total reflection from the interface between air ($n = 1$) and the lens edge ($n \sim 0$) coming from the medium with higher refractive index.

In our earlier work [14], we have developed a lithography technique which enables precise shape $d(r)$ control of the dielectric photoresist on gold film substrate. Unlike the traditional lithographic applications which aim for rectangular photoresist edges, we need to create a more gradual adiabatic edge profile. To produce gradual decrease of photoresist thickness (Shieply S1811 photoresist having refractive index n~1.5 was used for device fabrication) several methods have been used. Instead of contact printing (when mask is touching the substrate), we used soft contact mode (with the gap between the mask and the substrate). This allows for the gradient of exposure due to the diffraction at the edges, which leads to a gradual change of thickness of the developed photoresist. Underexposure and underdevelopment were also used to produce softer edges. Fabrication of arrays of devices has additional challenges: the lenses should be just touching each other. The fabrication procedures was tuned to avoid the device overlap, or significant separation.

Examples of so formed individual TO devices are presented in Figs. 1(b) and 2(a). As demonstrated by Fig. 1(c), we were able to fabricate photoresist patterns which almost ideally fit the modified Luneburg lens profile described by Eq.(5). Experimental images in Fig. 2 demonstrate measured performance of the individual Luneburg lenses. In these experiments a near-field scanning optical microscope (NSOM) fiber tip was brought in close proximity to the arrays of lithographically formed lenses and used as an illumination source. Almost diffraction-limited (~0.7$\lambda$) focusing of 515 nm light [14] emitted by the fiber tip (seen on the left) clearly demonstrates Luneburg lens-like focusing behavior of the fabricated devices for TM polarized light. Comparison of the theoretical and experimental images performed in Fig. 2(b) demonstrates excellent agreement between theory and experiment for both polarizations (artificial color scheme used to represent experimental images in Fig.2(b) has been chosen to better highlight this close match).

Let us now consider the concept of a TO waveguide based on a linear chain of Luneburg lenses, which has been developed in [1]. Its operation is obvious from Fig. 3(d), which shows a ray tracing simulation of the waveguide. A single Luneburg lens focuses incoming parallel rays onto a single spot located on its edge. Therefore, a linear set of Luneburg lenses placed next to each other is able to guide light while exhibiting periodic diffraction-limited foci, which are spaced at twice the lens diameter. The same result has been obtained in our numerical simulations of the straight Luneburg waveguide (Fig. 3(a,b)) performed using COMSOL Multiphysics using refractive index distribution corresponding to the experimental variation of waveguide thickness. However, if the Luneburg waveguide is curved, the ideal double periodicity appears to be broken, as shown in Fig. 3(c). Nevertheless, Mattheakis *et al.* predicted [1] that the Luneburg waveguides may be bent considerably without the loss of guiding. Our experimental results described below confirm these theoretical predictions.

Using the waveguide fabrication technique described above we were able to produce and study both straight and curved Luneburg waveguides, as demonstrated by Figs. 4-6. First, let us examine TM light propagation through a straight Luneburg waveguide. Microscopic images of the fabricated Luneburg waveguide taken at different magnifications are shown in Fig. 4 (a,c). This waveguide consists of a linear set of 1.5 μm diameter individual Luneburg lenses. The microscopic images of the waveguide taken while 488 nm light was coupled into the waveguide are shown in Fig. 4(b,d). The expected double periodicity of light distribution in the Luneburg waveguide is indicated by arrows in frames (c) and (d). This experiment clearly demonstrates a successful experimental realization of a Luneburg waveguide for the TM polarized light. Moreover, measurements of the polarization-dependent light propagation through the waveguide studied in Fig. 5 also confirm our theoretical model. As expected, TE polarized 488 nm light does not exhibit much propagation along the waveguide.

We have also studied propagation of TM light though curved Luneburg waveguides as illustrated in Fig. 6. As indicated by our theoretical simulations, the ideal double periodicity of light distribution appears to be broken in such waveguides. While some apparent double periodicity is indicated by arrows in Fig. 6(b), the cross-sectional analysis of the measured light distribution along the waveguide (Fig.6(c)) demonstrates that the double periodicity is generally broken. Nevertheless, the FFT analysis of the cross section (Fig. 6(d)) indicates that the double and the other even periods still dominate light distribution inside the curved waveguide, which confirms its Luneburg nature.

In conclusion, we have reported the first experimental realization of the TO-based Luneburg waveguides, which may be useful in sensing and nonlinear optics applications. We have studied wavelength and polarization dependent performance of the waveguides. Our technique opens up an additional ability to manipulate light on submicrometer scale.

**Acknowledgements**

This research was supported by the NSF grant DMR-1104676. We are grateful to J. Klupt and W. Zimmerman for experimental help.

**Figure Captions**

**Figure 1.** (a) Calculated effective refractive index for the TM and TE light as a function of thickness $d$ of the dielectric layer deposited onto the surface of ideal metal. The inset shows the dielectric waveguide geometry. (b) AFM image of the fabricated Luneburg lens. The inset shows 3D representations of the shape. (c) Measured photoresist thickness variation of the lens shown in (b) along the green line fitted to a modified Luneburg lens described by Eq. (5). The fit is shown in red. (d) Theoretical simulation of a waveguide-based Luneburg lens for TM polarized light using COMSOL Multiphysics. The lens diameter in the simulation is set to 1.

**Figure 2.** (a) Focusing behavior of the fabricated array of 6 μm diameter Luneburg lenses for TM polarized light. Additional white light illumination was used to highlight lens positions. (b) Digital zoom of the measured field distributions inside the Luneburg lens for TM and TE polarized light is shown next to theoretical simulations, which take into account real device shape. Artificial coloring scheme is used to differentiate between the signal and illuminating light.

**Figure 3.** Theoretical simulations of a Luneburg waveguide using COMSOL Multiphysics (a-c) and ray optics (d). Panel (a) shows effective refractive index distribution for TM light in a straight Luneburg waveguide, while panel (b) shows calculated energy density within the waveguide. Numerically calculated TM light propagation through a curved Luneburg waveguide (c) illustrates that double periodicity is typically broken within such a waveguide.

**Figure 4.** TM light propagation through experimentally fabricated straight Luneburg waveguide: (a,c) Microscopic images of the fabricated Luneburg waveguide taken at different magnifications. (b,d) Microscopic images of the same waveguide regions taken

while 488 nm light was coupled into the waveguide. Double periodicity of light distribution in the waveguide is indicated by arrows in frames (c) and (d).

**Figure 5.** Microscopic images of light propagation through the Luneburg waveguide taken using (a) TM and (b) TE polarized light.

**Figure 6.** Experimental analysis of TM light propagation through a curved Luneburg waveguide: (a) Microscopic image of the fabricated curved Luneburg waveguide. (b) Microscopic image of the same waveguide regions taken while 515 nm light was coupled into the waveguide. Double periodicity of light distribution in the waveguide is indicated by arrows. (c) Cross-section of image (b) measured along the waveguide demonstrates that similar to theoretical simulations in Fig. 3(c) double periodicity along the curved waveguide is broken. (d) FFT analysis of the cross section in (c) indicates that the double and the other even periods dominate light distribution inside the curved Luneburg waveguide.

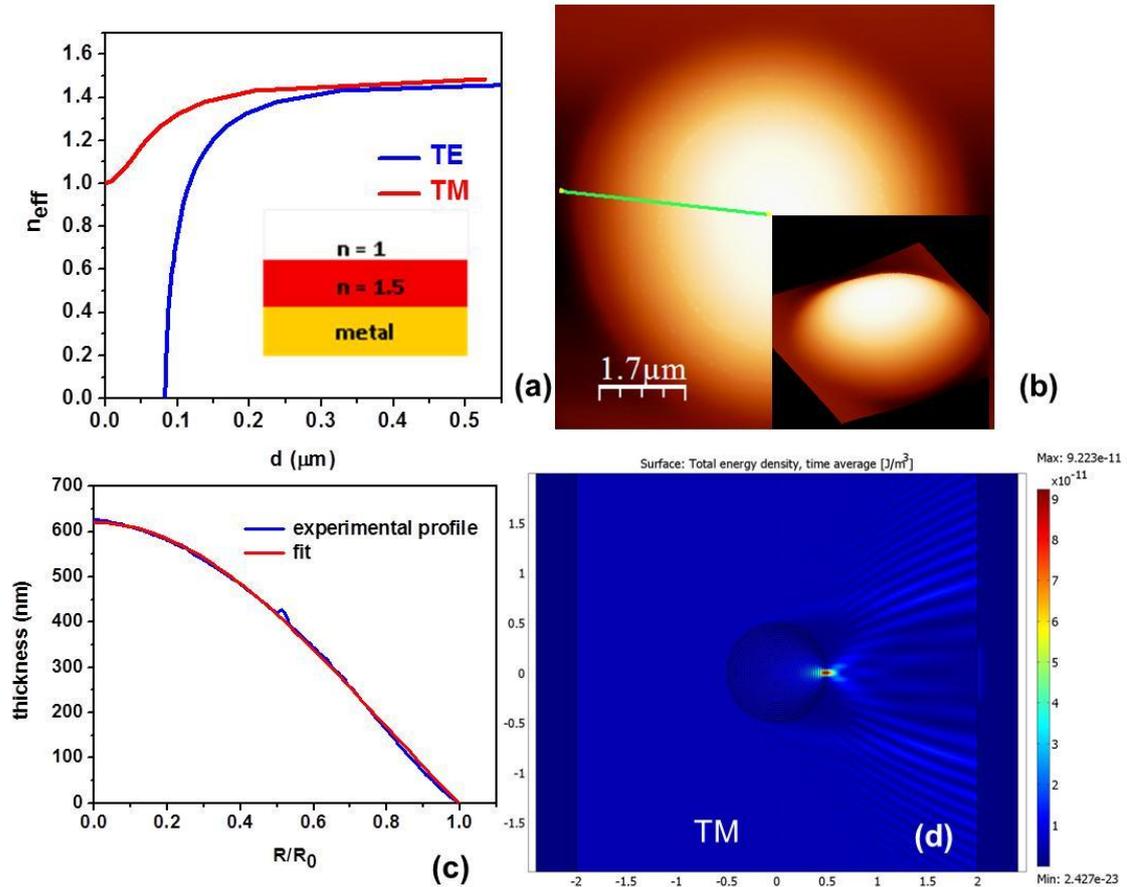

Fig.1

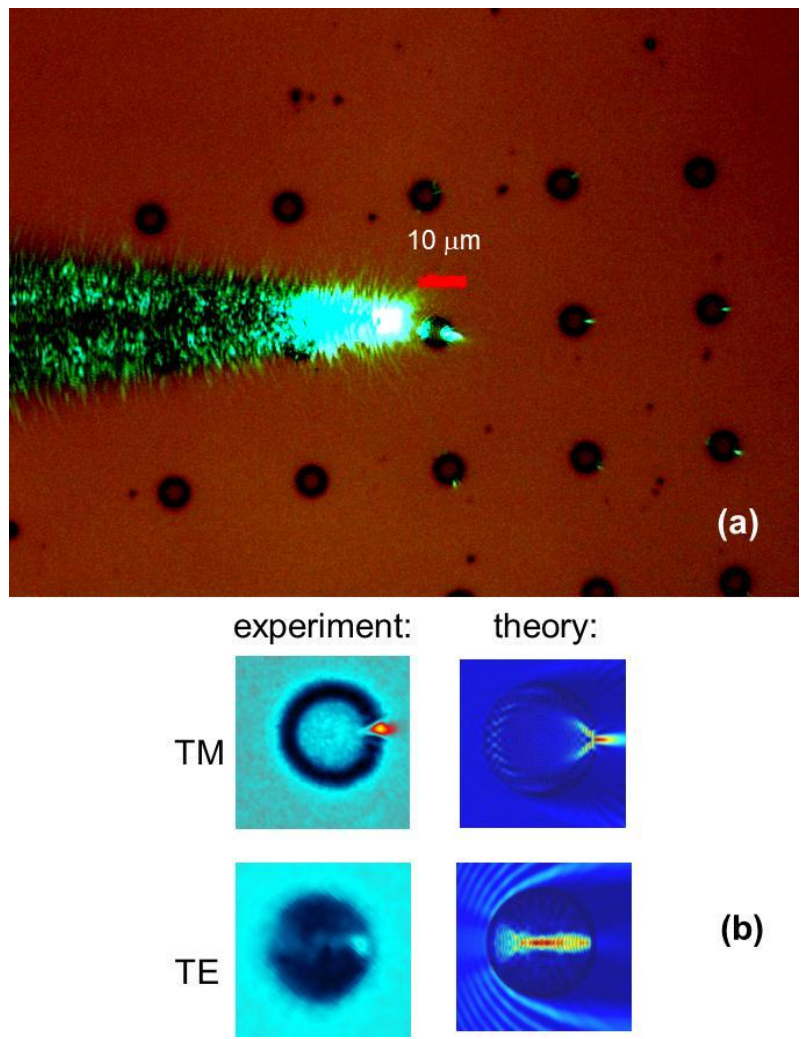

Fig. 2

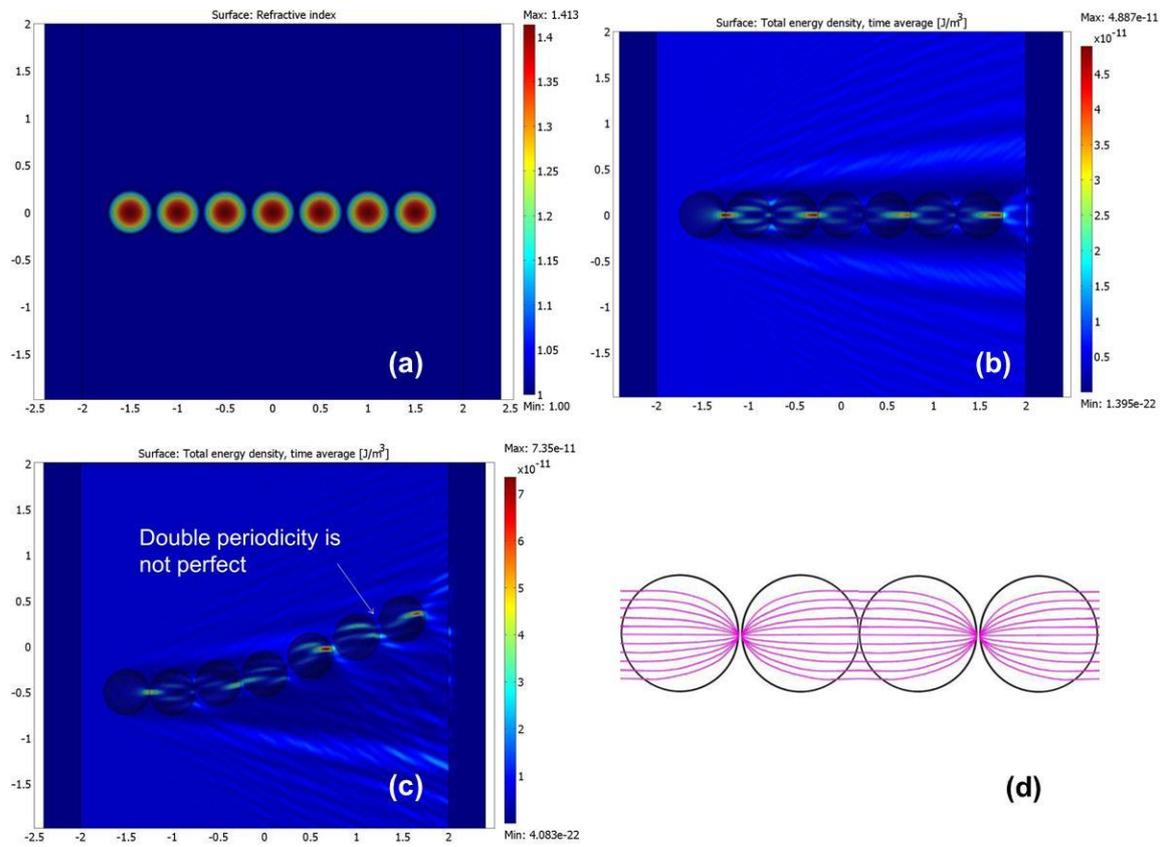

Fig. 3

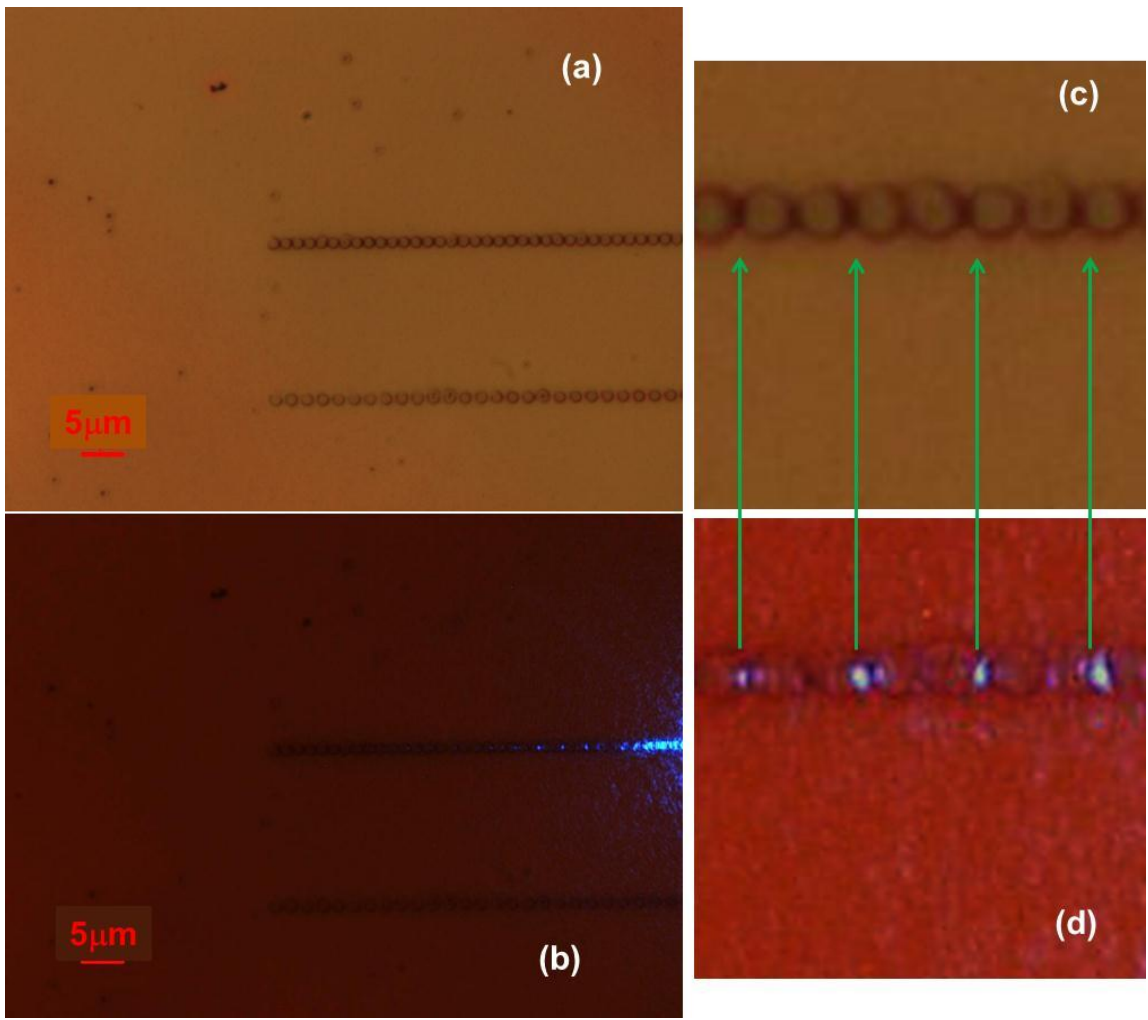

Fig. 4

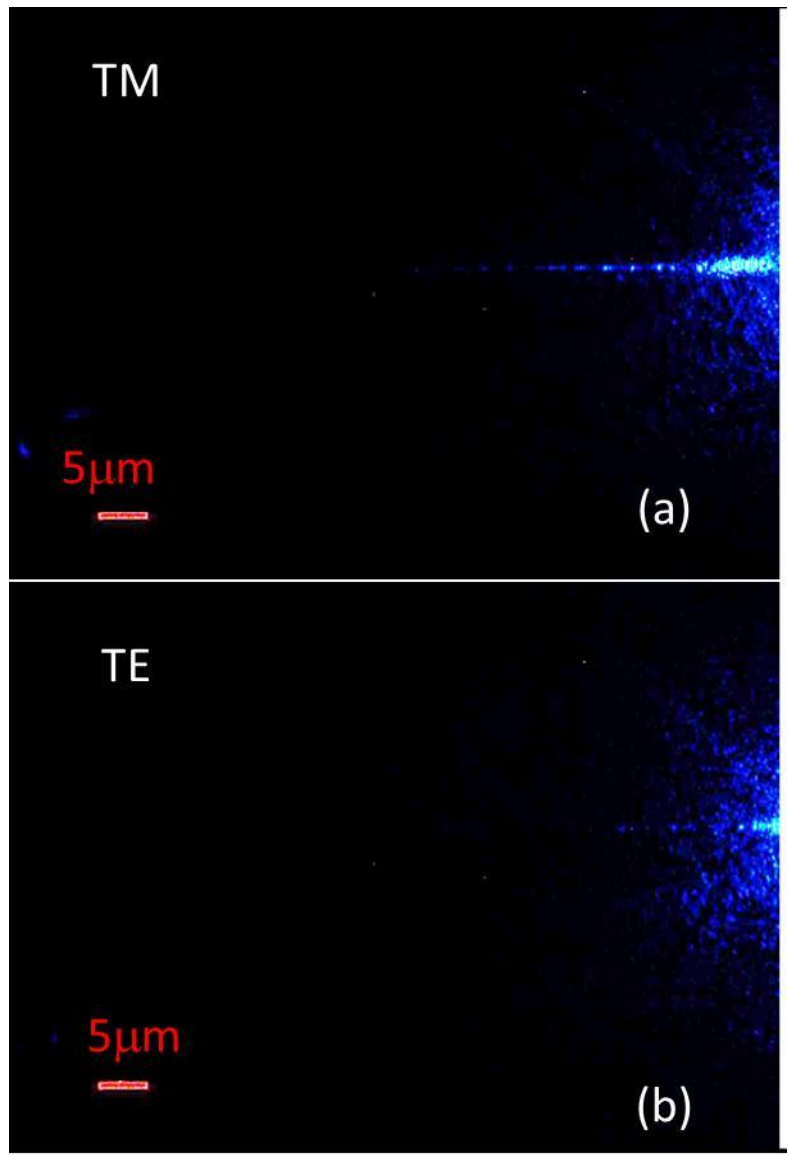

Fig. 5

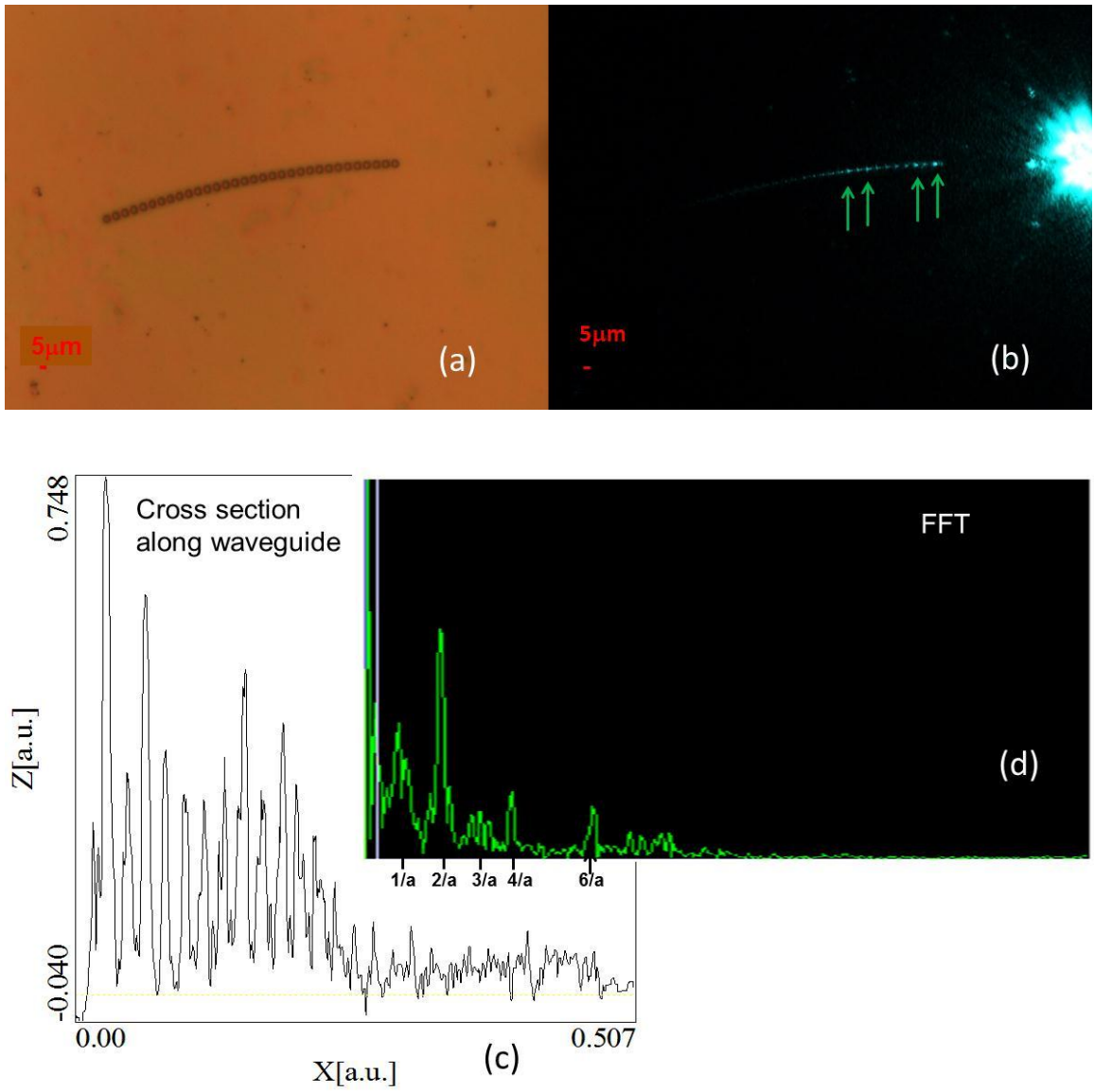

Fig. 6